\documentclass[journal]{IEEEtran}

\usepackage{epsfig}
\usepackage{authblk}
\usepackage{amsmath, amssymb, amsfonts} 
\usepackage{booktabs, graphicx, multirow, siunitx}
\usepackage{soul, subcaption, caption}
\usepackage{booktabs}
\usepackage{color, colortbl}
\definecolor{Highlight}{rgb}{1,0.9,0.7}
\usepackage[pagebackref=true]{hyperref}
\usepackage[pagewise]{lineno}
\usepackage[normalem]{ulem} 

\usepackage{pifont}
\usepackage{algorithmic}
\usepackage{textcomp}
\usepackage{xcolor}
\usepackage{csquotes}
\usepackage{tikz}
\usepackage{cleveref}
\usepackage[edges]{forest}
\usepackage{listings}
\usepackage{colortbl} 
\usepackage{float}
\usepackage{geometry} 
\usepackage{caption}  
\usepackage[linesnumbered,algoruled,boxed,lined]{algorithm2e}

\def\BibTeX{{\rm B\kern-.05em{\sc i\kern-.025em b}\kern-.08em T\kern-.1667em\lower.7ex\hbox{E}\kern-.125emX}}
\hypersetup{
    citecolor=blue,
    colorlinks=true,
    linkcolor=blue,
    filecolor=magenta,      
    urlcolor=cyan,
    pdftitle={Overleaf Example},
    pdfpagemode=FullScreen,
}

\begin{document}

\title{\LARGE Analysis of the Performance of the Matrix Multiplication Algorithm on the Cirrus Supercomputer
}

\author{Temitayo~Adefemi%
\thanks{Temitayo Adefemi is with the University of Edinburgh, Edinburgh, UK. (e-mail: T.M.Adefemi@sms.ed.ac.uk)}}

\maketitle
\setcounter{tocdepth}{3}

\begin{abstract}

Matrix multiplication is integral to various scientific and engineering disciplines, including machine learning, image processing, and gaming. With the increasing data volumes in areas like machine learning, the demand for efficient parallel processing of large matrices has grown significantly. This study explores the performance of both serial and parallel matrix multiplication on the Cirrus supercomputer at the University of Edinburgh. The results demonstrate the scalability and efficiency of these methods, providing insights for optimizing matrix multiplication in real-world applications.
\end{abstract}
\IEEEoverridecommandlockouts

\vspace{0.5cm}

\textbf{keywords: }{Matrix Multiplication, OpenMP, Parallelisation}

\section{Introduction}

Matrix multiplication is a fundamental linear algebra operation essential for numerous scientific applications. As these domains advance and the data scales represented in matrices grow, particularly in machine learning, the need for parallel tensor operations becomes increasingly critical. This research examines the performance of serial and parallel matrix multiplication on the Cirrus supercomputer at the University of Edinburgh.

\section{Literature Review}
Research on matrix multiplication algorithms showcases diverse methodologies to optimize computational efficiency, particularly in supercomputing. Traditional algorithms, such as Strassen's algorithm, have significantly reduced the computational complexity from $O(n^3)$. Yet, they encounter challenges when scaling to larger matrices, especially in parallel processing environments \cite{Mouhah2023}. These challenges include issues with load balancing, communication overhead, and memory access patterns, which become more pronounced as matrix sizes increase, particularly in distributed computing environments. With its tensor-based approach, the Coppersmith-Winograd algorithm offers further reductions in computational complexity, making it a promising candidate for large-scale applications. However, its practical implementation on supercomputers raises concerns regarding complexity and memory requirements, which can limit its applicability in real-world scenarios \cite{Zhang2024}.

\vspace*{0.3cm}

Supercomputing environments introduce additional challenges and opportunities for matrix multiplication algorithms. Parallel Matrix Multiplication (PMM) algorithms, such as those explored by Liao et al. \cite{Liao2021}, are designed to leverage supercomputers' high processing power and vast memory bandwidth. These algorithms optimize data communication and power distribution across multiple nodes, significantly enhancing the efficiency of matrix operations. Efficiently managing communication between nodes is crucial, as it can mitigate the bottlenecks typically associated with large-scale matrix computations.

\vspace*{0.3cm}

Furthermore, specialized algorithms like the Jacobi-Davidson method exploit supercomputers' unique capabilities to solve significant sparse matrix problems with high accuracy. This method exemplifies how advanced hardware can push the boundaries of existing algorithms, allowing for the computation of complex eigenvalue problems that would be otherwise infeasible on standard computing platforms \cite{Kinateder2001}.

\vspace*{0.3cm}

Recent research has also focused on hybrid approaches that combine the strengths of classical algorithms with modern innovations in hardware and parallel processing. For instance, the TileSpMV algorithm for GPUs, as developed by Niu et al. \cite{Niu2021}, showcases how a tiled approach to sparse matrix-vector multiplication can be optimized for GPUs, significantly enhancing performance over traditional methods. Similarly, Baransel et al. \cite{Baransel2014} proposed a new parallel matrix multiplication algorithm for wormhole-routed 2D/3D torus networks, which improves scalability and reduces communication overhead in massively parallel supercomputers.

\vspace*{0.3cm}

The ongoing development of these algorithms is driven by the need to handle ever-increasing data sizes and the growing complexity of scientific and engineering applications. As matrix multiplication remains a cornerstone of computational science, future research will likely focus on developing hybrid approaches that combine the strengths of classical and modern algorithms with the processing power of supercomputers to achieve unprecedented efficiency and scalability \cite{Liu2023}. This will involve refining existing algorithms and exploring new architectures and paradigms that can better exploit the capabilities of next-generation supercomputers, as highlighted in studies such as those by Xu et al. \cite{Xu2021} and Johnson \cite{Johnson2020}.

\vspace*{0.3cm}

In conclusion, while significant progress has been made in optimizing matrix multiplication algorithms for supercomputing environments, challenges remain, particularly in scaling these algorithms for ever-larger matrices. The continued evolution of both algorithmic strategies and computing hardware will be essential to meeting the demands of future scientific and industrial applications.

\section{Methodology for Parallelising Matrix Multiplication}

Matrix multiplication is an operation that combines two matrices to produce a new matrix. To multiply matrices 

Let \( \mathbf{A} \) be an \( m \times n \) matrix and \( \mathbf{B} \) be an \( n \times p \) matrix. The resulting matrix \( \mathbf{C} \) will be \( m \times p \).

\[
\mathbf{C}[i][j] = \sum_{k=1}^n \mathbf{A}[i][k] \times \mathbf{B}[k][j]
\]

For each element \( \mathbf{C}[i][j] \):

\begin{itemize}
    \item Take the i-th row of A:
    (A[i][1], A[i][2], . . . , A[i][n])
    \item Take the j-th column of B: (B[1][j],
    B[2][j], . . . , B[n][j])
    \item Perform the dot product of these two vectors:
    \begin{align*}
    \mathbf{C}[i][j] = & \, \mathbf{A}[i][1] \times \mathbf{B}[1][j] + \mathbf{A}[i][2] \times \mathbf{B}[2][j] \\
    & + \ldots + \mathbf{A}[i][n] \times \mathbf{B}[n][j]
    \end{align*}
\end{itemize}

\vspace{0.3cm}

\renewcommand{\thefigure}{\arabic{figure}} 

\begin{figure}[h]
    \centering
    \includegraphics[width=0.45\textwidth]{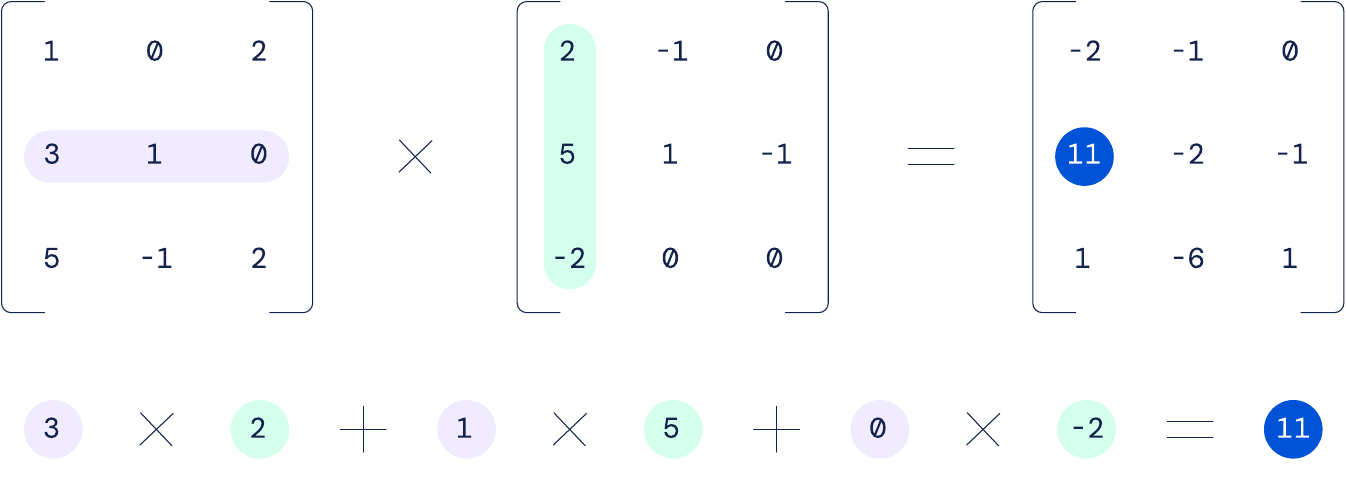} 
    \caption{Matrix Multiplication}
    \label{fig:matmul}
\end{figure}

Due to the nature of the algorithm, the mode and method of parallelization matter, and the language used is crucial. In this experiment, the programming language utilized is C, which stores data in row-major order in memory. Other languages like Fortran, which uses column-major order, might yield different memory access and computation speed efficiencies.

The matrix multiplication sequence can be illustrated as follows:

This operation is repeated for all \( i = 1 \) to \( m \) and \( j = 1 \) to \( p \) to fill the entire \( \mathbf{C} \) matrix.

\vspace{0.3cm}

\paragraph{Column Prefetching}

The row-major storage model in C is well-suited for accessing rows of \( \mathbf{A} \), potentially enhancing cache performance for these operations. However, accessing columns of \( \mathbf{B} \) in this storage scheme is less efficient due to stridden memory access patterns. Although optimizations, such as pre-storing the columns of \( \mathbf{B} \) in an array before performing multiplication, could be considered, they generally yield minimal improvements. This inefficiency arises because the columns still need to be accessed, and the additional overhead of managing these elements in an array may slightly degrade performance.

\vspace{0.3cm}
\paragraph{Cache Tiling}

To address the inefficiencies discussed, this report will explore the use of cache tiling techniques. Cache tiling improves row and column access by dividing the matrix into smaller blocks. This optimization allows for more efficient use of cache memory, as data can be reused from the cache rather than being repeatedly accessed from main memory. The effectiveness of this optimization is expected to scale with the number of matrix multiplications computed using the same matrices and the size of the matrix blocks. Cache tiling can significantly enhance performance by optimizing access patterns and data reuse, particularly in scenarios with high computation reuse.

\subsection{Parallelisation Paradigms}

This section explores various paradigms for effectively parallelizing matrix multiplication. We will compare threads versus processes, each offering distinct advantages and drawbacks regarding utilization. These points will be discussed in detail to understand the options available in sufficient detail.

\subsubsection{Threads}

\begin{figure}[H]
    \centering
    \includegraphics[width=0.4\textwidth]{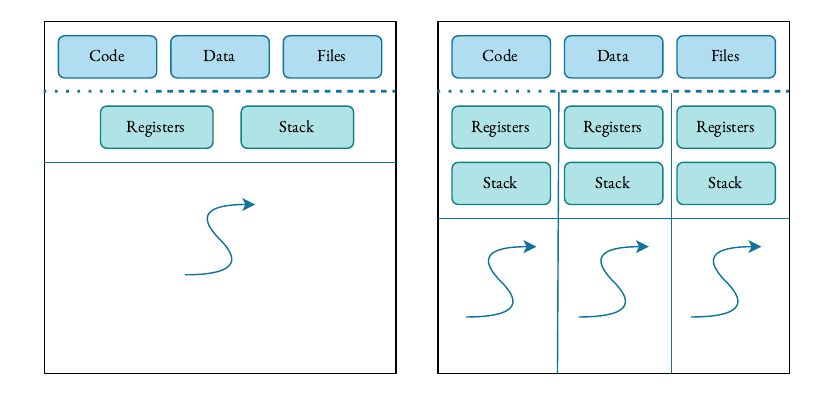} 
    \caption{Single vs Multi-threaded programs}
    \label{fig:matmul}
\end{figure}

Utilizing threads is the easiest and most common method of parallelizing this algorithm. OpenMP provides practical pragma directives that allow parallelization with just one line of code above the algorithm. By leveraging OpenMP's tried and tested methods, we can avoid reinventing the wheel and rewriting our thread parallelization algorithm from scratch. This approach minimizes the risk of introducing race conditions and synchronization issues.

\vspace{0.3cm}

Moreover, thanks to the shared memory model of threads, we don't need to worry about communication to share variables between threads. This is particularly advantageous as it eliminates a potential bottleneck when scaling up to larger matrices.

\subsubsection{Processes}

\begin{figure}[h]
    \centering
    \includegraphics[width=0.3\textwidth]{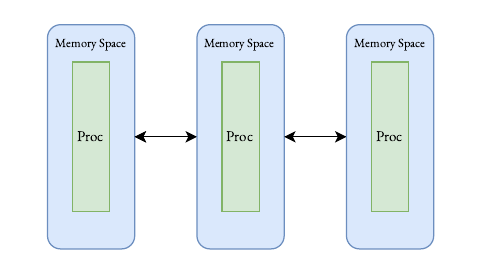} 
    \caption{Illustration of Processes}
    \label{fig:matmul}
\end{figure}

Parallelizing a matrix multiplication algorithm using processes is not as straightforward as parallelizing using threads. Matrix multiplication may be better suited for processes than problems such as a lattice calculation or a cellular automaton problem.
Matrix multiplication and cellular automata pose distinct challenges in parallel computing using MPI. Cellular automata require each process to access border cells, typically managed through halo exchange. Conversely, matrix multiplication involves accessing data in a linear but sporadic fashion. Efficient distributed matrix multiplication algorithms like Cannon's algorithm or SUMMA (Scalable Universal Matrix Multiplication Algorithm) reduce communication overhead and balance the workload effectively. These algorithms are particularly advantageous for large matrices but might be excessive for smaller ones. There are other proposed solutions in research, such as the 2.5D Matrix Multiplication and a method called Tesseract. According to the calculations by [], The communication needed for the Cannon's Algorithm is 31.5 times the communication necessary for the Tesseract, and the communication needed for the 2.5D algorithm is 3.75 times the communication required by the Tesseract.

\begin{figure}[h]
    \centering
    \includegraphics[width=0.4\textwidth]{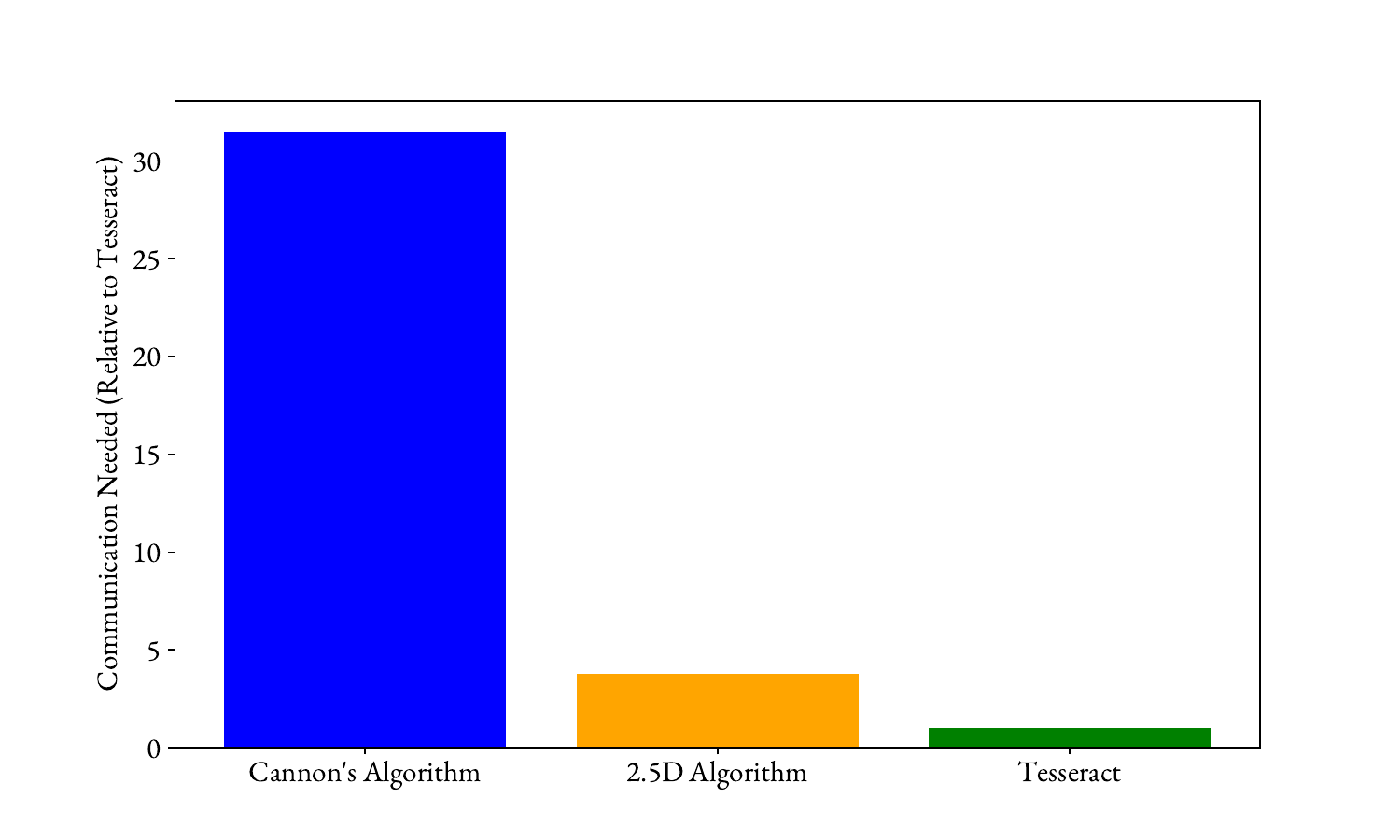} 
    \caption{Communication Needed for Different Algorithms Relative to Tesseract}
    \label{fig:matmul}
\end{figure}

\vspace{0.3cm}

Implementing these algorithms can be more complex than a single-process, multi-threaded solution. For small matrices or when quick implementation is critical, multi-threading on a single process might be more appropriate. Conversely, an MPI-based solution could offer better performance for large matrices where scalability across multiple nodes is needed.

\vspace{0.7cm}

The decision between using MPI and multi-threading depends on several factors: the size of the matrix, available hardware, scalability needs, and the time available for development. Additionally, the performance of a multi-threaded or MPI program can be influenced by the software ecosystem, from hardware capabilities to compiler efficiency. However, parallelization using OpenMP for multi-threading may be the preferred approach for straightforward implementations where these factors are less critical.

\section{Experimental Dataset}

We employ the Box-Muller transform to generate matrices of random numbers for our matrix multiplication experiments. This method enables the creation of normally distributed random numbers, which helps simulate real-world data and ensure a broad range of values in our matrices. This is to provide balance to our matrices so the distribution of large and small numbers can be balanced to give a nuanced view of each matrix efficiency proportioned to the size of the matrix, the parallelization methodology, patterns, and optimizations.

\vspace{0.3cm}

The Box-Muller transform algorithm is as follows:
\begin{enumerate}
    \item Generate two independent uniform random numbers \( U1 \) and \( U2 \) in the range (0, 1).
    \item Calculate two normally distributed random numbers \( Z0 \) and \( Z1 \):
    \[
    Z0 = \sqrt{-2 \ln(U1)} \cos(2\pi U2)
    \]
    \[
    Z1 = \sqrt{-2 \ln(U1)} \sin(2\pi U2)
    \]
\end{enumerate}

We implement this in C as follows, shown in Figure~\ref{fig:boxmuller}:

\begin{figure}[h]
\centering
\begin{lstlisting}[language=C, basicstyle=\scriptsize\ttfamily, breaklines=true]
#include <math.h>
#include <stdlib.h>

void box_muller(double *z0, double *z1) {
    double u1, u2;
    u1 = (double)rand() / RAND_MAX;
    u2 = (double)rand() / RAND_MAX;
    *z0 = sqrt(-2.0 * log(u1)) * cos(2.0 * M_PI * u2);
    *z1 = sqrt(-2.0 * log(u1)) * sin(2.0 * M_PI * u2);
}
\end{lstlisting}
\caption{Pseudocode for the Box-Muller transform in C.}
\label{fig:boxmuller}
\end{figure}

\subsection*{Experimental Matrix Sizes and Dimensions}

We use a range of matrix sizes and dimensions that grow in scale to derive essential insights from our parallelization experiments. This approach allows us to observe how the performance of our algorithms changes with increasing problem size and to identify any potential bottlenecks or scalability issues.

We propose the following matrix sizes for our experiments:
\begin{itemize}
    \item Small matrices: 32x32, 64x64, 128x128
    \item Medium matrices: 256x256, 512x512, 1024x1024
\end{itemize}

\vspace{0.3cm}

Using the Box-Muller transform, we will generate matrices for each matrix size and perform matrix multiplication using our serial and parallelized algorithms. We will measure and record the performance of the matrices on the Cirrus Supercomputer.

\section{Experimental Environment}

To gain a comprehensive understanding of the programs, we conducted a rigorous statistical analysis with 30 preliminary tests to account for data variability. A power analysis was performed using the following formula to determine the necessary sample size for detecting a medium effect size of 0.5 with 80\% power (0.8) and a 5\% significance level (0.05):

\[
n = 2 \left( \frac{Z_{1-\frac{\alpha}{2}} + Z_{1-\beta}}{0.5} \right)^2 \sigma^2
\]

Where \( Z_{1-\frac{\alpha}{2}} \) and \( Z_{1-\beta} \) are the critical values from the normal distribution for the specified significance and power levels, respectively, and \( \sigma^2 \) represents the variance of the data.

\vspace{0.3cm}

Given the data variability, our preliminary analysis indicated that 15 iterations were sufficient to produce significant results.

\vspace{0.3cm}

The serial version of the program was executed 15 times for each matrix size, while the parallel version was also run 15 times for each thread count. These tests were applied across various optimization methodologies, with constant refining of the algorithm to evaluate shifts in performance and potential benefits of using Intel or GCC compilers. This approach facilitated a detailed discussion on thread distribution and the impact of optimization on performance. The testing process was automated through robust bash scripts to manage unexpected results and address potential issues.

\vspace{0.3cm}

All experiments were conducted on the Cirrus HPC system at the University of Edinburgh, which features 283 compute nodes equipped with 2.1 GHz, 18-core Intel Xeon E5-2695 processors, each with 256 GB of memory and three levels of cache:

\begin{itemize}
    \item L1 Cache: 32 KiB (per core)
    \item L2 Cache: 256 KiB (per core)
    \item L3 Cache: 45 MiB (shared)
\end{itemize}

The GPU nodes house two 2.4 GHz, 20-core Intel Xeon Gold 6248 (Skylake) series processors, supporting two hardware threads per core by default, and provide:
\begin{itemize}
    \item 8 GPU accelerators
    \item 80 CPU cores
    \item Three levels of cache:
    \begin{itemize}
        \item L1 Cache: 32 KiB (per core)
        \item L2 Cache: 1 MiB (per core)
        \item L3 Cache: 27.5 MiB (shared)
    \end{itemize}
\end{itemize}
Compiler settings for the Intel experiments used Intel 20.4 compilers with the \texttt{-O3 -qopenmp} options, while GNU compiler experiments were conducted using GNU 10.2 with the -O3 -fopenmp flags for GNU compiler experiments.

\section{Results \& Analysis}

We conducted experiments using a basic matrix multiplication algorithm. Our main function accepted three parameters:

\begin{itemize}
    \item Size of the first matrix
    \item Size of the second matrix
    \item Size of the resulting matrix (which stores the calculated values)
\end{itemize}

We used the \texttt{atoi} function to extract these values from the environment—the \texttt{malloc} function allocated memory for each row and column of each matrix. Values were assigned to the two input matrices using the Box-Muller transform algorithm.

We then performed a straightforward matrix multiplication without any optimizations. The GCC and Intel compilers measured the calculation time across various matrix sizes. For each compiler, we ran two sets of experiments:
\begin{itemize}
    \item Using no optimization flags
    \item Using the \texttt{-O3} optimization flag
\end{itemize}

The results of these experiments are presented below.

\begin{figure}[h]
    \centering
    \includegraphics[width=0.5\textwidth]{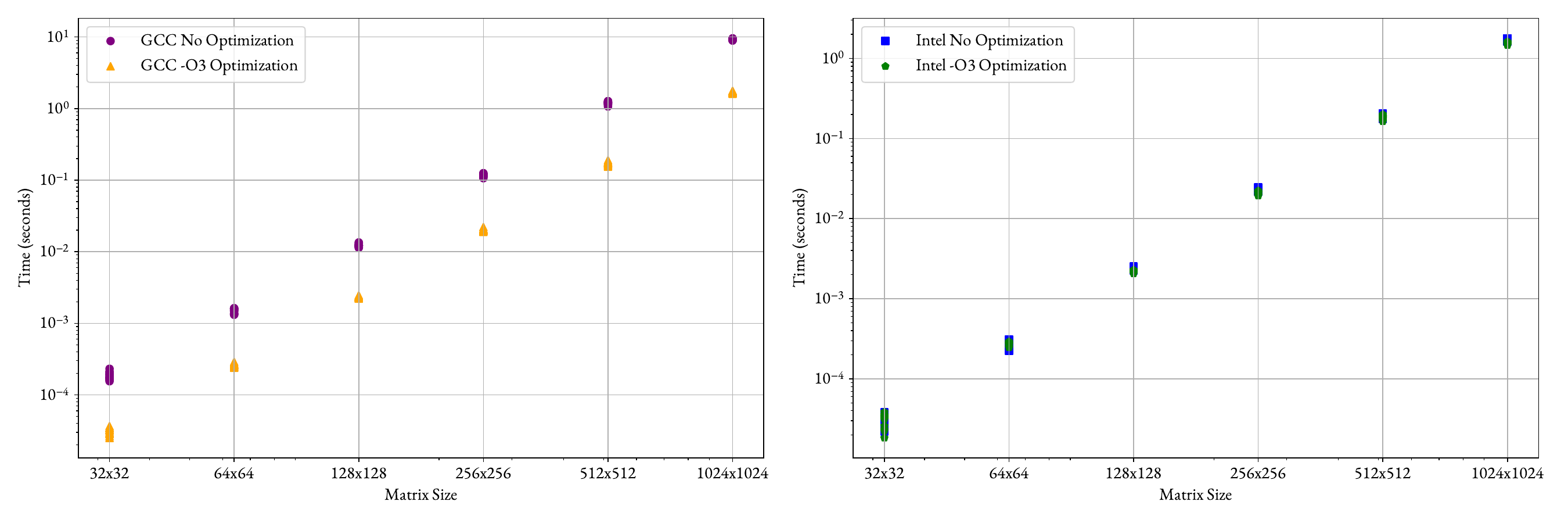} 
    \caption{Graph showing Comparative Performance of GCC and Intel Compilers with and without Optimization (-O3)}
    \label{fig:matmul}
\end{figure}

\newcolumntype{C}[1]{>{\centering\arraybackslash}m{#1}}
\definecolor{Highlight}{rgb}{0.8,0.9,1}  

\noindent 
\begin{minipage}[t]{0.49\textwidth}
\centering
\begin{tabular}{|C{1.6cm}|C{2.2cm}|C{2.2cm}|}
\hline
\rowcolor{Highlight}
\multicolumn{3}{|c|}{\bfseries\small GCC Compiler} \\
\hline
\textbf{\small Matrix Size} & \textbf{\small Optimization} & \textbf{\small Avg Time (sec)} \\
\hline
32x32 & No Opt & 0.00018 \\
\hline
32x32 & -O3 & \cellcolor{Highlight}0.00003 \\
\hline
64x64 & No Opt & 0.00145 \\
\hline
64x64 & -O3 & \cellcolor{Highlight}0.00026 \\
\hline
128x128 & No Opt & 0.01214 \\
\hline
128x128 & -O3 & \cellcolor{Highlight}0.00228 \\
\hline
256x256 & No Opt & 0.11887 \\
\hline
256x256 & -O3 & \cellcolor{Highlight}0.02028 \\
\hline
512x512 & No Opt & 1.19296 \\
\hline
512x512 & -O3 & \cellcolor{Highlight}0.17037 \\
\hline
1024x1024 & No Opt & 9.12568 \\
\hline
1024x1024 & -O3 & \cellcolor{Highlight}1.69891 \\
\hline
\end{tabular}
\captionof{table}{Performance of GCC Compiler with and without Optimization}
\end{minipage}%
\hfill 
\begin{minipage}[t]{0.49\textwidth}
\centering
\begin{tabular}{|C{1.6cm}|C{2.2cm}|C{2.2cm}|}
\hline
\rowcolor{Highlight}
\multicolumn{3}{|c|}{\bfseries\small Intel Compiler} \\
\hline
\textbf{\small Matrix Size} & \textbf{\small Optimization} & \textbf{\small Avg Time (sec)} \\
\hline
32x32 & No Opt & 0.00003 \\
\hline
32x32 & -O3 & 0.00003 \\
\hline
64x64 & No Opt & 0.00027 \\
\hline
64x64 & -O3 & 0.00027 \\
\hline
128x128 & No Opt & 0.00237 \\
\hline
128x128 & -O3 & 0.00215 \\
\hline
256x256 & No Opt & 0.02313 \\
\hline
256x256 & -O3 & \cellcolor{Highlight}0.02081 \\
\hline
512x512 & No Opt & 0.19141 \\
\hline
512x512 & -O3 & \cellcolor{Highlight}0.18232 \\
\hline
1024x1024 & No Opt & 1.71092 \\
\hline
1024x1024 & -O3 & \cellcolor{Highlight}1.55254 \\
\hline
\end{tabular}
\captionof{table}{Performance of Intel Compiler with and without Optimization}
\end{minipage}

\vspace{0.7cm}

Our data shows that matrix multiplication time increases cubically with input size, matching naive matrix multiplication's expected $O(n^3)$ complexity. We observed that optimizations have a more significant impact when using the GCC compiler, especially for larger matrices. Several factors can explain this:

\paragraph{Cache Efficiency}
The processor's multi-level cache allows faster data access as calculations are repeated across rows and columns. The specific cache hierarchy of our 2.1 GHz, 18-core Intel Xeon E5-2695 processors (32 KiB L1 per core, 256 KiB L2 per core, and 45 MiB shared L3) plays a crucial role:
\begin{itemize}
  \item Small matrices ($\leq 64 \times 64$) likely fit entirely in L1 cache, leading to minimal performance differences between compilers.
  \item Medium matrices ($64 \times 64$ to $180 \times 180$) utilize L2 cache extensively, where compiler optimizations for cache usage become critical.
  \item Large matrices ($180 \times 180$ to $2400 \times 2400$) rely heavily on L3 cache, where efficient prefetching and cache utilization strategies significantly impact performance.
  \item Huge matrices ($> 2400 \times 2400$) exceed L3 cache capacity, requiring optimizations for managing cache misses and efficient main memory access.
\end{itemize}

\paragraph{Pipelining}
As matrix size increases, the processor can better utilize pipelining, overlapping multiple instructions for improved efficiency. This becomes particularly important for larger matrices that can't fit entirely in lower-level caches.

\paragraph{Compiler Optimizations}
Various low-level optimization techniques become more effective with larger computations. These likely include:
\begin{itemize}
  \item Efficient blocking/tiling strategies to maximize cache usage at each level.
  \item Intelligent prefetching to minimize cache misses, especially for L3 and main memory access.
  \item Loop unrolling and vectorization to maximize the use of each cache line fetched.
\end{itemize}

GCC demonstrates superior optimization, particularly for larger matrices, due to its effective use of cache and pipelining. Given the algorithm's simplicity, the \texttt{-O3} flag enables aggressive optimizations like vectorization and prefetching, which are safe to use. GCC's performance advantage suggests it might be implementing these strategies more effectively for this specific processor architecture, especially in leveraging the large shared L3 cache for more extensive matrix computations.

\vspace{0.3cm}

The Intel compiler shows less dramatic optimization effects. While some performance gains are seen with larger matrices using the \texttt{-O3} flag, smaller matrices show minimal improvement. This suggests that the Intel compiler may already be well-optimized for the processor's architecture at lower computational loads but might not be as effective in utilizing the large L3 cache for this particular workload.

\vspace{0.3cm}

Another important finding from the results shows that the -03 optimization flag is particularly effective in optimizing the code utilizing the GCC compiler but not particularly effective when using the Intel Compiler when running on the Intel Xeon processor. The -O3 flag enables aggressive optimizations, including loop unrolling, vectorization, inlining, and other advanced techniques. However, the way these optimizations are applied varies between compilers. GCC (GNU Compiler Collection) and the Intel Compiler (ICC) are optimized differently; GCC is a general-purpose compiler with optimizations that aim to perform well across a wide range of hardware, including Intel Xeon processors, but without deep hardware-specific tuning. In contrast, the Intel Compiler is designed with detailed knowledge of Intel hardware. It applies optimizations that are more fine-tuned to the architectural features of Intel processors, such as the Xeon series. Consequently, while GCC’s -O3 optimizations may yield significant performance improvements due to broad-spectrum optimizations, they may not align as closely with the specific microarchitectural features of the Xeon processors as those generated by the Intel Compiler. The Intel Compiler may employ different or more conservative strategies under -O3 based on its understanding of the hardware, leading to less dramatic improvements than GCC.

\vspace{0.3cm}

As matrix size increases, both compilers can perform better through increased vectorization, pipelining, and a more extensive search space for optimizations due to more permutations. To investigate these differences further, one could measure cache hit/miss rates for different matrix sizes, analyze the generated assembly code for blocking and prefetching implementations, and experiment with matrix sizes targeting specific cache level boundaries.

\begin{figure}[h]
    \centering
    \includegraphics[width=0.5\textwidth]{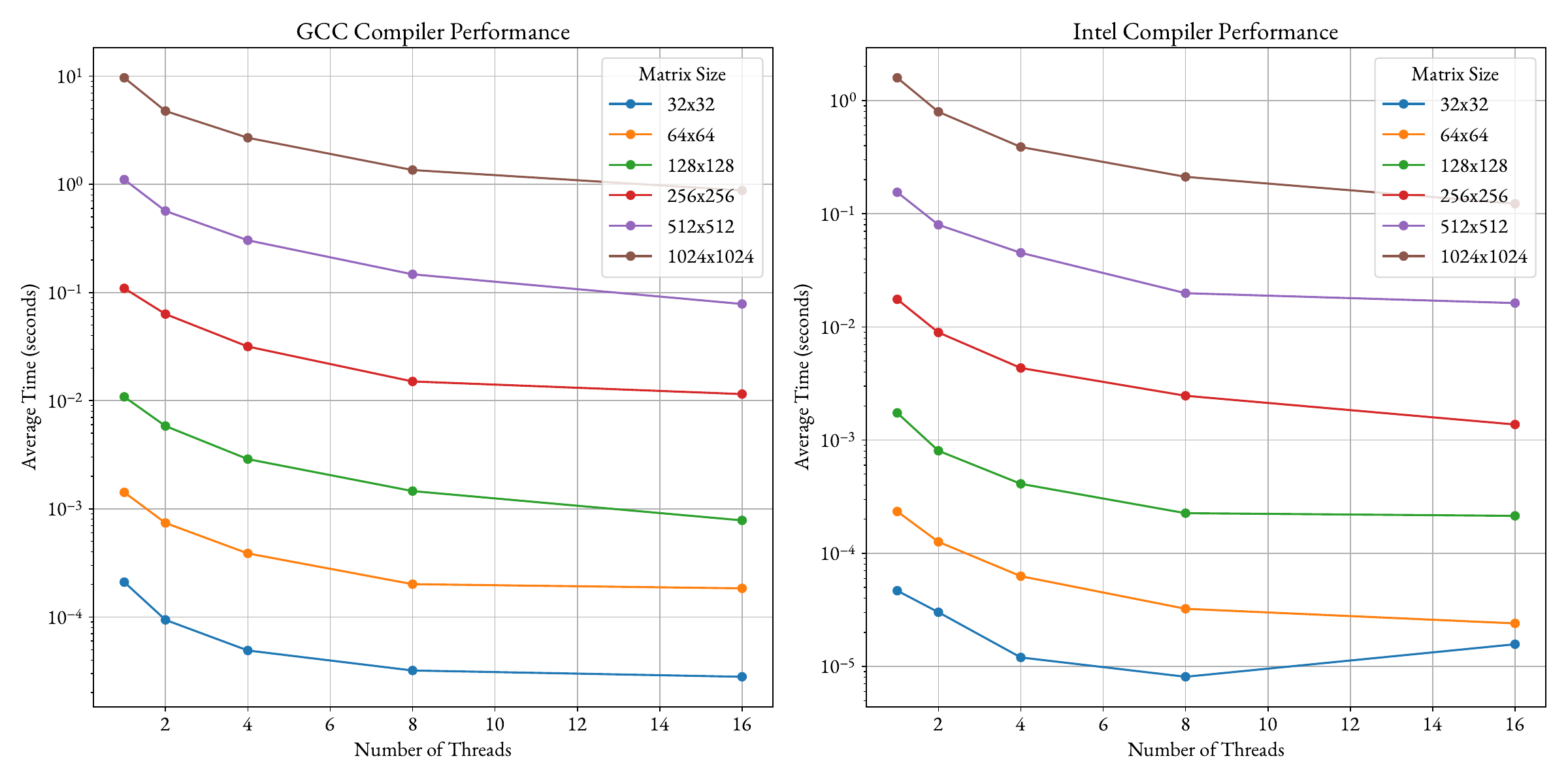} 
    \caption{Graph showing Comparative Performance of GCC and Intel Compilers across thread counts}
    \label{fig:matmul}
\end{figure}

The comparison between GCC and Intel compilers reveals significant differences in performance, mainly as thread counts and matrix sizes vary. For both compilers, increasing the thread count leads to a substantial reduction in computation time. However, the Intel compiler consistently outperforms GCC, especially in handling smaller matrices and higher thread counts. For instance, with a 1024x1024 matrix, the GCC computation time decreases from 9.67 seconds with one thread to 0.88 seconds with 16 threads, showing a roughly 10x improvement. Conversely, Intel achieves a more impressive reduction from 1.59 seconds to 0.12 seconds over the same range. This indicates that Intel’s optimizations are particularly effective in multi-threading scenarios.
Additionally, while both compilers show exponential increases in computation time with larger matrices, Intel maintains an edge, particularly with smaller matrices like 32x32, where its performance is markedly superior. For example, with a 32x32 matrix and 16 threads, Intel's time is 0.0000157 seconds, compared to GCC’s 0.000028 seconds, highlighting Intel's advantage in efficiently managing smaller tasks. However, the performance gap narrows as matrix size increases, suggesting GCC might be more competitive in more significant matrix scenarios. While Intel is generally the superior compiler, especially for small matrices and high thread counts, GCC can still be viable for larger matrices where the performance disparity is less pronounced.

\vspace{0.3cm}

\paragraph{CPU}
This phenomenon can be attributed to several factors. Firstly, larger matrices often shift the computational bottleneck from the CPU to the memory subsystem, where the relative advantages of Intel's optimizations are less pronounced. While not as optimized for CPU-intensive tasks, GCC handles memory-bound tasks with comparable efficiency, leading to a smaller performance gap. Secondly, more extensive matrix operations may involve more cache misses and memory accesses, where the advantage of specialized CPU instructions (often leveraged by Intel) is reduced. Lastly, Intel's more sophisticated thread management might give it an edge with small matrices in high-thread-count scenarios. Still, as matrix sizes grow, the complexity of memory access patterns can equalize performance. Thus, GCC becomes more competitive in large matrix operations due to its less aggressive but broader optimization approach, making it a viable alternative in scenarios where the matrix size mitigates the specific advantages of Intel's optimizations.

\paragraph{Hyperthreading}

\begin{figure}[h!]
    \centering
    \includegraphics[width=0.5\textwidth]{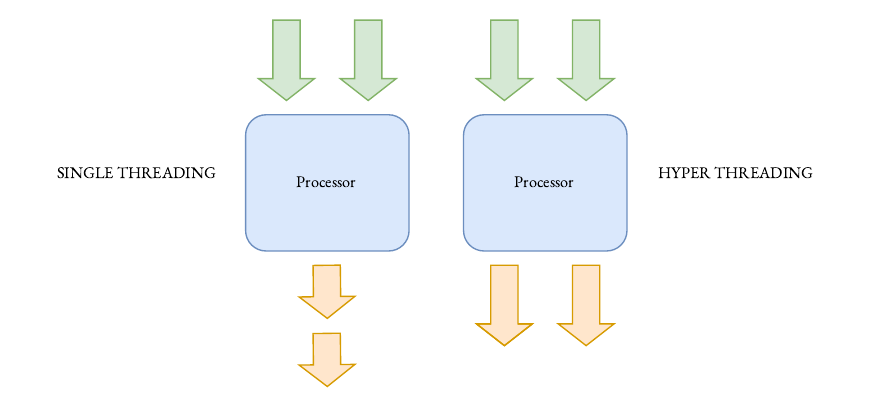} 
    \caption{Illustration of Hyperthreading}
    \label{fig:matmul}
\end{figure}

When considering the Intel Xeon processor architecture, an increase in thread count leverages the processor's multi-core and hyper-threading capabilities to boost performance in matrix multiplication. Intel Xeon processors, designed for high-performance computing, feature multiple cores, each capable of handling various threads simultaneously through Intel's Hyper-Threading Technology (HTT). This allows a single physical core to execute two threads concurrently, doubling the number of threads the processor can handle, assuming sufficient resources.

As thread count increases on an Intel Xeon processor, each core must manage more execution threads, which can enhance performance significantly in tasks like matrix multiplication, where operations can be divided among threads. For instance, in a matrix multiplication task, other threads can process different matrix portions simultaneously, reducing overall computation time. On an Xeon processor with 16 cores and HTT enabled, you could theoretically run up to 32 threads concurrently; this implies that although the Intel Xeon only has 16 counts, there is still potential for significant gains if we increase the thread count for this experiment.

However, this increase in thread count would also cause challenges to underlying architecture in several ways. First, the shared resources, such as caches (L1, L2, and L3), are put under increased pressure as more threads vie for space, potentially leading to cache thrashing where the effective use of the cache is reduced. Memory bandwidth also becomes a critical factor; as more threads are introduced, more memory requests are generated, which can saturate the available bandwidth, leading to performance bottlenecks. The Xeon's QuickPath Interconnect (QPI), designed to enhance data transfer between cores, must efficiently manage this increased traffic to avoid latency issues.

\section{Conclusion}

The comprehensive analysis conducted in this report highlights the critical role of parallelization and optimization techniques in enhancing the performance of matrix multiplication on high-performance computing platforms. The experiments demonstrate that multi-threading with OpenMP offers significant speedups for moderate-sized matrices. At the same time, MPI-based multi-processing is more effective for large-scale problems requiring scalability across multiple nodes. Compiler optimizations, particularly with GCC, yield substantial performance improvements, especially for larger matrices that challenge cache and memory subsystems. The findings suggest that parallelization strategy and compiler optimizations should be tailored to the specific matrix size and hardware architecture to maximize computational efficiency. Future research could explore further optimizations and alternative algorithms to push the boundaries of matrix multiplication performance in increasingly complex and data-intensive applications.

\end{document}